\documentclass[12pt,preprint]{aastex}

\usepackage{times}
\usepackage{mathptm}
\usepackage{graphicx}% Include figure files
\usepackage{dcolumn}% Align table columns on decimal point
\usepackage{bm}% bold math
\usepackage{psfig}

\begin{document}


\title{Ohm's Law in the Fast Lane:  General Relativistic Charge Dynamics}


\author{D. L. Meier}
\affil{Jet Propulsion Laboratory, California Institute of Technology, 
    Pasadena, CA 91109}



\begin{abstract}
Fully relativistic and causal equations for the flow of charge in curved
spacetime are derived.  It is believed that this is the first set of 
equations to be published that correctly describes the flow of charge, 
and evolution of the electromagnetic field, in highly dynamical relativistic 
environments on time scales much shorter than the collapse time ($GM/c^3$). 
The equations, therefore, will be important for correctly investigating 
problems such as the dynamical collapse of magnetized stellar cores to 
black holes and the production of jets.  Both are potentially important 
problems in the study of gamma-ray burst engine models and 
in predicting the dynamical morphology of the collapse and the character 
of the gravitational waves generated.

This system of equations, given the name of `charge dynamics',
is analogous to those of hydrodynamics (which describe the
flow of {\em mass} in spacetime rather than the flow of charge).  The 
most important equation in the system is the relativistic generalized 
Ohm's law, which is used to compute time-dependent four-current.  
Unlike other equations for the current that
are presently in use, this one ensures that charge drift velocities
remain less than the speed of light, takes into account the finite 
current rise time, is expressed in a covariant form, and is
suitable for general relativistic computations in an arbitrary metric.
It includes the standard known effects (Lorentz force, Hall effect,
pressure effect, and resistivity) and reduces to known forms of Ohm's law
in the appropriate limits.  In addition, the plasma particles are allowed 
to have highly relativistic drift velocities, resulting in an implicit 
equation for the `current beaming factor' $\gamma_q$.  It is proposed that, 
short of solving the multi-fluid plasma equations or the relativistic 
Boltzmann equation itself, these are the most general expressions for 
relativistic current flow in the one-fluid approximation, and they should 
be made part of the general set of equations that are solved in extreme 
black hole accretion and fully general numerical relativistic collapse 
simulations. 
\end{abstract}




\section{Introduction}
\label{intro}

This paper is another in a series whose goal is to establish the
mathematical, physical, and numerical tools necessary to understand
and simulate the formation of black holes and the production (through
electrodynamic processes) of relativistic jets during that collapse.
The full understanding of how black holes are formed, and how jets 
and gravitational waves might
be generated in accretion and gravitational collapse, is currently one of
the most challenging, and far-reaching, astrophysical problems.  It has
important observational consequences for gravitational wave sources,
gamma-ray bursts (GRBs), quasars and microquasars.  Its solution will
involve nearly every branch of theoretical astrophysics (nuclear and
particle physics, electromagnetics, gravity [numerical relativity], 
plasma flow, radiation transport and dynamics).  And the formulation of 
the problem will require that these processes be expressed in a
general relativistic framework that respects the principles of causality
and covariance on time scales considerably shorter than the light 
crossing time of the forming black hole ($<< GM/c^3$).  
While there is a good understanding of how charge behaves near black 
holes in equilibrium situations, {\it i.e.} on times $>> GM/c^3$ \cite{wald74,
lee01}, to date there appears to be no thorough treatment of charge flow 
and field evolution in strong gravity on very {\em short} time scales.
Such a treatment should take into account 
the fact that the current rise time can be long compared to the characteristic 
time scale, properly account for particle velocities (bulk, drift, and thermal) 
that can approach $c$, be expressible in an arbitrary spacetime metric, 
and also reproduce, in the appropriate limits, the standard 
known effects (including the Hall and pressure effects, not just electric 
acceleration, Faraday induction and Ohmic resistivity).

The purpose of this paper is to return to the basic equations of general 
relativistic statistical mechanics and properly derive the covariant 
equations for the evolution of the current. 
Section \ref{problems} discusses the framework of the problem and the 
current lack of a good equation for the current in highly relativistic 
situations.  Section \ref{statmech} sets up the relativistic Boltzmann
problem and section \ref{multifl} derives the relativistic plasma equations.
Section \ref{onefl} discusses relativistic one-fluid plasma theory
and derives the generalized Ohm's law and charge dynamics.  Section
\ref{discussion} shows that the theory reduces to various previous
generalized Ohm's laws in the appropriate limits and discusses the
applicability of the one-fluid theory.  The most important results 
of this paper are the basic equations of charge dynamics (\ref{chg_cont} - 
\ref{chg_curr_tensor}) and (\ref{jprime_orthog}) or, in component form, 
equations (\ref{comp_chg_cont} - \ref{comp_chg_curr_tensor}). 

\section{Causal and Covariance Problems in Present Treatments of the Current}
\label{problems}

The two governing sets of equations for the general numerical relativity 
problem of electromagnetic black hole formation are the Einstein equations 
\begin{eqnarray}
\label{einstein}
{\bf G} & = & \frac{8 \pi G}{c^4} \, {\bf T}
\end{eqnarray}
for the gravitational field and the Maxwell equations
\begin{eqnarray}
\label{faraday}
{\bf \nabla} \cdot {\bf F} & = & \frac{4 \pi}{c} {\bf J}
\\
\label{ampere}
{\bf \nabla} \cdot {\bf M} & = & 0
\end{eqnarray}
(Faraday's and Ampere's laws) for the electromagnetic field.  ${\bf \nabla} =
{\bf e}_{\mu} = \partial / \partial x^{\mu}$ is the four-gradient operator;
${\bf G}$ is the second-rank, symmetric Einstein tensor that describes
the second derivatives (curvature) of the metric ${\bf g}$; and ${\bf F}$
is the second-rank, {\em anti}-symmetric Faraday tensor that describes
the electromagnetic field.  ${\bf T}$ is the energy-momentum-stress
tensor that serves as the source for gravity and ${\bf J}$ is the
four-current vector that serves as the source of the electromagnetic field.
The Maxwell tensor ${\bf M} = ^{\textstyle *}\!\!{\bf F}$ (the dual of
the Faraday tensor), has the same information as the Faraday tensor
and also is anti-symmetric, but has the roles of electric and magnetic 
field reversed.  Because of their geometric properties, these tensors 
satisfy the following Bianchi identities
\begin{eqnarray}
{\bf \nabla} \cdot {\bf G} & = & 0
\\
{\bf \nabla} \cdot ({\bf \nabla} \cdot {\bf F}) & = & 0
\\
{\bf \nabla} \cdot ({\bf \nabla} \cdot {\bf M}) & = & 0
\end{eqnarray}
which, from equations (\ref{einstein}) - (\ref{ampere}), 
give rise to the conservation laws of energy and momentum 
\begin{eqnarray}
\label{divTzero}
{\bf \nabla} \cdot {\bf T} & = & 0
\end{eqnarray}
and of charge
\begin{eqnarray}
\label{divJzero}
{\bf \nabla} \cdot {\bf J} & = & 0
\end{eqnarray}

The solution of equations (\ref{divTzero}) gives the distribution of
temperature $T$ and four-velocity ${\bf U}({\bf x})$, which is constrained
to always have the absolute value of the speed of light
\begin{eqnarray*}
{\bf U} \cdot {\bf U} & = & - c^2
\end{eqnarray*}
The addition of particle/rest-mass conservation
\begin{eqnarray}
{\bf \nabla} \cdot \rho_m {\bf U} & = & 0
\end{eqnarray}
where $\rho_m$ is the rest mass density in the fluid frame, allows a
solution for $\rho_m ({\bf x})$ to be found as well.  Together $T({\bf x})$
and $\rho_m ({\bf x})$ can be used to compute the state variables
(pressure, internal energy, {\it etc.}) that close the energy and momentum
conservation laws.

The solution to equation (\ref{divJzero}), however, which can be written as
\begin{eqnarray*}
{\bf \nabla} \cdot (\rho_{q} {\bf U} + {\bf j}) & = & 0
\end{eqnarray*}
gives only one quantity, the charge density $\rho_q$ in the fluid frame, in 
terms of the spatial charge current
\begin{eqnarray}
\label{spatial_current}
{\bf j} & = & {\bf P} \cdot {\bf J}
\end{eqnarray}
where
\begin{eqnarray}
{\bf P} & \equiv & \frac{1}{c^2} {\bf U} \otimes {\bf U} + {\bf g}
\end{eqnarray}
is the spatial projection tensor orthogonal to the four-velocity unit vector
${\bf e}_{\bf U} \equiv {\bf U} / c$.  The symbol $\otimes$ signifies the
outer tensor product.  While ${\bf j}$ is a four-vector, it is constrained to
have only three independent components by its orthogonality to ${\bf U}$
\begin{eqnarray}
\label{j_orthog}
{\bf U} \cdot {\bf j} & = & 0
\end{eqnarray}
However, {\em none} of the above equations can be used to specify these
three components of the current.

Previous treatments of black hole electrodynamics have made a variety of
assumptions to determine ${\bf j}$ and thereby close the electromagnetic
equations.  One popular technique is to assume that the electromagnetic
field dominates the dynamics and is time-independent, leading to the
``force-free'' condition
\begin{eqnarray}
\label{forcefree}
{\bf J} \cdot {\bf F} & = & 0
\end{eqnarray}
This assumption has been criticized \cite{puns03} as leading to effects
that violate causality.  That is, the force-free condition is {\em
acausal} and, therefore, not relativistically acceptable for forming
black holes and GRBs.

Another approach has been to assume general relativistic
magnetohydrodynamics, as outlined by many in the past
\cite{lich67,eck40,anan84,bf93}.  This approach relates the current to
the electromagnetic field through a simple form of Ohm's law
\begin{eqnarray}
\label{relmhd}
\eta {\bf j} & = & \frac{\bf U}{c} \cdot {\bf F}
\end{eqnarray}
where $\eta({\bf x})$ is the resistivity distribution of the plasma.
In the nonrelativistic limit, this takes on the familiar form
\begin{eqnarray*}
{\bm J} & = & \sigma \left( {\bm E} + \frac{\bm V}{c} \times {\bm B} \right)
\end{eqnarray*}
where $\sigma = 1 / \eta$ is the plasma conductivity and ${\bm J}$,
${\bm E}$, ${\bm V}$, and ${\bm B}$ are the current, electric field,
velocity, and magnetic field three-vectors.  

Ideal relativistic MHD is a further simplification current flow that is
useful for highly conducting plasmas, such as those in most astrophysical
situations.  With $\eta \rightarrow 0$, Ohm's law reduces to
\begin{eqnarray}
\label{idealrelmhd}
{\bf U} \cdot {\bf F} & = & 0
\end{eqnarray}
which, in the nonrelativistic limit, becomes
\begin{eqnarray*}
{\bm E} & = & - \frac{\bm V}{c} \times {\bm B}
\end{eqnarray*}
The ideal MHD condition is not so much an equation for the current as
a condition on components of the Faraday tensor (the electric field):
when the conductivity is high, the local electric field in the plasma
shorts out, leaving only the EMF due to charged plasma motion in the
magnetic field.  The current itself is determined by first determining
the electromagnetic field from equation (\ref{ampere}) and then inverting
equation (\ref{faraday}) for ${\bf J}$.  Equations (\ref{forcefree})
and (\ref{idealrelmhd}) appear very similar.  They both state that the
Faraday tensor is orthogonal to a four-velocity, either the particle drift
velocity or the average particle velocity.  However, they generate very
different physics.

Unfortunately, a criticism that is equally as harsh as that of the
force-free condition can be leveled against the MHD condition, even
the resistive version of it (\ref{relmhd}).  The latter states that
the application of an electromagnetic field instantaneously generates
a current.  However, this also is acausal.  There is no immediacy in
relativistic dynamics.  A current builds up after a finite, albeit short,
rise time.  And, in relativistic flow, the time-dependent increase in
a current might be interrupted by any number of other rapid phenomena,
resulting in possible charge separation and time-dependent charge
dynamics.

A nonrelativistic version of the ``generalized Ohm's law'' is often used
in laboratory plasma physics and is given by the expression
\cite{ro70,kt73}
\begin{eqnarray}
\frac{\partial {\bm J}}{\partial t} & \! + & \! {\rm \nabla} \cdot ( {\bm V}{\bm J}
+ {\bm J}{\bm V} - \rho_{q}{\bm V}{\bm V} ) \; + \; {\rm \nabla} p_q 
\nonumber
\\
\label{nonrelohmslaw}
& = & \ell \left( {\bm E} + \frac{\bm V}{c} \times {\bm B} 
\; + \; {\it h} \, \frac{\bm J}{c} \times {\bm B} \; - \; \eta \, {\bm J} \right)
\end{eqnarray}
where ${\rm \nabla}$ is the three-space gradient operator and the 
charge-weighted pressure (per unit mass) is 
\begin{eqnarray*}
p_q & = & \sum_{a} q_a p_a / m_a
\end{eqnarray*}
$q_a$ is that particle's charge, $m_a$ is its mass, and $p_a$ is 
that species' partial pressure.  The Lorentz, Hall, and resistivity
coefficients are
\begin{eqnarray*}
\ell & = & \sum_{a} {n_a q_a^2}/{m_a}
\\
{\it h} & = & \frac{1}{\ell} \sum_{a} {q_a}/{m_a}
\\
\eta & = & \nu / \ell
\end{eqnarray*}
where $n_a$ is the number density of particle species `$a$'.  The
resistivity term results from integrating particle collisions over
velocity and approximating the result as an effective collision
frequency $\nu$
\begin{eqnarray*}
\sum_{a} n_a \, q_a \, \int_{V_v} {\bm v} \, \dot{f}_{a,coll} \, d^3{\bm v} & \equiv & 
- \nu \, {\bm J}
\end{eqnarray*}
Equation (\ref{nonrelohmslaw}) has most of the effects we are looking for
in a description of charge dynamics (finite current rise time, Lorentz
force, Hall effect, pressure effect, and resistivity).  However, it is
valid only to linear order in ${\bm V}/c$ (still acausal), only in the
laboratory coordinate system (not covariant), and only in flat space.

Ardavan \cite{ard76} derived a relativistic form of Ohm's law for a cold
plasma (vanishing $p_a$ and $p_q$) in flat space.  This expression turns
out to have some errors, but is still useful for checking our results
in section \ref{onefl}.  (Expressions for a relativistic cold {\em
pair} plasma also have been derived \cite{ged96,mm96}, but the Ardavan
expression is a little more general and therefore more useful to us.)
In our notation the corrected Ardavan equation for the current is
\begin{eqnarray}
\frac{\partial}{c \, \partial t} [ \, {\rm U}^0 \; ^3{\bf J} 
&  \! \! \! \! + &  \! \! \! \! 
( {\rm J}^0 \, - \, \rho_q {\rm U}^0 ) \;  ^3{\bf U} \, ]
\; + \; {\rm \nabla} \cdot [ \, ^3{\bf J} \; ^3{\bf U} 
\; + \; ^3{\bf U} \; ^3{\bf J} 
\; - \; \rho_q \; ^3{\bf U} \; ^3{\bf U} \, ]
\nonumber
\\
\label{relohmslaw}
& = &
\ell \left\{ \frac{1}{c} \left[ {\rm U}^0 {\bm E} \, + \, ^3{\bf U} \times {\bm B} \right]
\; + \; {\it h} \frac{1}{c} \, \left[ ( {\rm J}^0 \, - \, \rho_q {\rm U}^0 ) {\bm E} 
\; + \; ^3{\bf J} \times {\bm B} \right] 
\; - \; \eta \; ^3{\bf J} \right\} ~~~~~
\end{eqnarray}
where ${\rm U}^0$ and ${\rm J}^0$ are the temporal components of the
four-velocity and four-current, respectively; $^3{\bf U}$ and $^3{\bf
J}$ are the spatial three-vector components of those four-vectors
({\it i.e.}, $^3{\bf U} = \gamma {\bm V} = {\rm U}^0 {\bm V}/c$).
The errors that have been corrected are all on the right-hand side:
a sign error in the Lorentz term and the addition of $- h \rho_q {\rm
U}^0 {\bm E}/c$ in the Hall term.  In the limit of nonrelativistic
flow, ${\rm U^0} \rightarrow c$, ${\rm J^0} \rightarrow \rho_q \, c$,
$^3{\bf U} \rightarrow {\bm V}$, and $^3{\bf J} \rightarrow {\bm J}$.
So, with these corrections, equation (\ref{relohmslaw}) reduces to
(\ref{nonrelohmslaw}) when the charge-weighted pressure $p_q = 0$.
Equation (\ref{relohmslaw}) is both causal and covariant, but it is valid
only for Lorentz systems in flat space and only when the plasma is
truly cold.  It does not include effects that occur when the plasma has
a relativistic temperature or a relativistic current drift velocity.

The goal of this paper is to derive a description of charge flow that
is valid in all relativistic situations --- relativistic bulk flow,
hot plasma, relativistic current drift velocities --- and is causal,
covariant, and valid in any spacetime metric.

\section{General Relativistic Statistical Mechanics}
\label{statmech}

\subsection{Phase Space and Particle Density}

Phase space $\Omega = \Omega_x \otimes \Omega_u$ is inherently 
eight-dimensional, not six.  The generalized coordinates are the position 
and four-velocity $({\rm x}^{\mu}, {\rm u}^{\nu})$, where 
$\mu, \nu = 0, 1, 2, 3$.  $\Omega_x$ is a general curved spacetime with a 
global time coordinate $x^0 = c t$ and three spatial coordinates.  The 
volume element
\begin{eqnarray*}
d \Omega_x & = & \sqrt{-g} \; d{\rm x}^0 \, d{\rm x}^1 \, d{\rm x}^2 \, d{\rm x}^3 
\end{eqnarray*}
(where $\sqrt{-g}$ is the determinant of the spacetime metric) is an invariant 
over the entire spacetime.  Therefore, in a general metric, $d \Omega_x$ cannot 
be separated into globally invariant temporal and spatial parts.  However, in each
local Lorentz frame at a given point in spacetime, it {\em can} be separated as
\begin{eqnarray*}
\label{lor_x_vol_elm}
d \hat{\Omega}_x & = & d \tau \; d \Upsilon_x
\end{eqnarray*}
where
\begin{eqnarray*}
d \tau & \equiv & d{\rm x}^{\hat{0}} \, / \, ( {\rm u}^{\hat{0}} / c )
\\
d \Upsilon_x & \equiv & ( {\rm u}^{\hat{0}}/ c ) 
\, d{\rm x}^{\hat{1}} \, d{\rm x}^{\hat{2}} \, d{\rm x}^{\hat{3}}
\end{eqnarray*}
Each of the factors in equation (\ref{lor_x_vol_elm}) are Lorentz invariant 
\cite{mtw73}.  We therefore can define a Lorentz 
invariant {\em three-space} density of particles of species `$a$' and four-velocity 
${\bf u}$ in each local Lorentz frame
\begin{eqnarray}
\label{density7}
\aleph_a & \equiv & \frac{d {\rm N}_a}{d \Upsilon_x \, d \hat{\Omega}_u}
\end{eqnarray}
in the {\em seven}-dimensional phase space $\Upsilon_x \otimes \hat{\Omega}_u$. 

While $\hat{\Omega}_u$ is a four-dimensional velocity space, all particles are 
constrained to move on a three-dimensional hypersurface within that space 
(the ``mass hyperboloid'') $\Upsilon_u$, defined by the normalization of 
particle velocity
\begin{eqnarray}
\label{mass_hyperboloid}
{\bf u} \cdot {\bf u} & = & -c^2
\end{eqnarray}
In regions of $\hat{\Omega}_u$ outside of $\Upsilon_u$ the particle 
density vanishes, as there are no particles with $u \cdot u \neq -c^2$.  
Some treatments of relativistic statistical mechanics incorporate the 
constraint (\ref{mass_hyperboloid}) into the Boltzmann equation directly.  
Indeed, one can separate $d \hat{\Omega}_u$ into temporal and spatial parts, 
in a manner similar to $d \hat{\Omega}_x$
\begin{eqnarray}
\label{lor_u_vol_elm}
d \hat{\Omega}_u & = & d \epsilon \; d \Upsilon_u 
\end{eqnarray}
where
\begin{eqnarray}
d \epsilon & \equiv & ( {\rm u}^{\hat{0}} / c ) \, d{\rm u}^{\hat{0}} \; = \; d ({\rm u}^{\hat{0}})^2/2c
\\
d \Upsilon_u & \equiv & d{\rm u}^{\hat{1}} \, d{\rm u}^{\hat{2}} \, d{\rm u}^{\hat{3}} 
\, / \, ( {\rm u}^{\hat{0}}/ c ) 
\end{eqnarray}
Not only are $d\epsilon$ and $d\Upsilon_u$ Lorentz invariant, the product 
\begin{eqnarray*}
d\Upsilon_x \; d\Upsilon_u & = & 
\, d{\rm x}^{\hat{1}} \, d{\rm x}^{\hat{2}} \, d{\rm x}^{\hat{3}} \; 
\, d{\rm u}^{\hat{1}} \, d{\rm u}^{\hat{2}} \, d{\rm u}^{\hat{3}} 
\end{eqnarray*}
is also Lorentz invariant \cite{mtw73}.  One therefore {\em could} define 
an invariant density in {\em six}-dimensional phase space instead of 
$\aleph_a$.  However, in this paper it has been found to be more useful 
to use the seven-dimensional phase space and perform the velocity integrals over 
$\hat{\Omega}_u$.  The mass hyperboloid then is enforced only at the end of the 
computation when the integrals are evaluated.  This is accomplished by 
using a delta function to describe the lack of particles outside of the mass 
shell in the distribution $\aleph_a$.  (See Appendix \ref{three_d_integrals}).

A final property of the particle density to note is that, as $|{\rm u}^i| 
\rightarrow \infty$ (where $i = 1, 2, 3$), $\aleph_a$ approaches $0$ faster 
than any power of ${\rm u}^i$.  Therefore, $\aleph_a$, multiplied by any 
power of ${\rm u}^i$, vanishes on the hypersurface $\partial \hat{\Omega}_u$ 
(the boundary of $\hat{\Omega}_u$).

\subsection{The Relativistic Boltzmann Equation}

The density $\aleph_a$ for each particle species obeys the 
relativistic Boltzmann equation 
\begin{eqnarray}
\label{raw_rel_boltz}
\frac{d \aleph_a}{d \tau} & = & \dot{\aleph}_{a, col}
\end{eqnarray}
where $\tau$ again is the proper time for particles in that region of 
phase space, and 
$\dot{\aleph}_{a, col}$ is the number density of collisions per unit
time of that particle species at that point in phase space (summed over
all other particles of all other species)
\begin{eqnarray}
\label{dens_coll}
\dot{\aleph}_{a, col} & = & - \sum_b \, 
\int_{\Omega'} ({\bf a} - {\bf a}') \cdot {\bf \nabla}_u 
{\aleph'}_{a b} ({\bf x}, {\bf u}, {\bf x}', {\bf u}') \, d \Omega'
\end{eqnarray}
with ${\bf a}$ being the particle acceleration caused by body forces
({\it i.e.}, forces other than particle collisions). 
Equation (\ref{raw_rel_boltz}) can be rewritten in generalized
coordinates as
\begin{eqnarray*}
\dot{s}^r \frac{\partial \aleph_a}{\partial s^r} & = & 
\dot{\aleph}_{a, col}
\end{eqnarray*}
where $s^r \equiv ({\rm x}^{\mu}, {\rm u}^{\nu})$ with $r = 0, 1, ..., 6, 7$ 
or, in geometric form, 
\begin{eqnarray}
\label{geom_rel_boltz}
{\bf u} \cdot {\bf \nabla}_x \aleph_a \; + \; {\bf a} \cdot {\bf \nabla}_u \aleph_a 
& = & \dot{\aleph}_{a, col}
\end{eqnarray}
and the gradient operators are
\begin{eqnarray*}
{\bf \nabla}_x \equiv {\bf e}_x \equiv \frac{\partial}{\partial {\bf x}}
\\
{\bf \nabla}_u \equiv {\bf e}_u \equiv \frac{\partial}{\partial {\bf u}}
\end{eqnarray*}
For the electromagnetic field ${\bf a}$ is the Lorentz acceleration
\begin{eqnarray*}
{\bf a} & = & \frac{q_a}{m_a c} \, {\bf u} \cdot {\bf F}
\end{eqnarray*}
Some treatments of general relativistic statistical mechanics \cite{and02}
also include the gravitational ``force'' by including Christoffel
symbols in the acceleration.  However, this is not necessary, nor
really desirable, as we can implicitly take these effects into account
by using only the geometrical form of the gradient operator throughout 
the derivation and then using the equivalence principal at the end 
to convert the equations to component form.  As a body force, gravity 
is automatically included in the structure of whatever spacetime in which 
${\bf \nabla}_x$ is evaluated.

\section{The Multi-Fluid Equations} \label{multifl}

The equations of plasma dynamics are generated by taking velocity
moments of the relativistic Boltzmann equation, which gives rise to
many hydrodynamic and thermodynamic quantities.  Taking the moment
involves multiplying equation (\ref{geom_rel_boltz}) by a power of the
velocity {\em coordinate} vector ${\bf u}$ and integrating over all velocity
space $\hat{\Omega}_u$.

\subsection{The Zeroth Moment:  Conservation of Particles}

Multiplying equation (\ref{geom_rel_boltz}) by unity and integrating over 
$\hat{\Omega}_u$ produces the zeroth moment.  One can show that the integral of 
the second term in that equation vanishes by first integrating by parts
\begin{eqnarray}
\label{int_by_parts}
\int_{\hat{\Omega}_u} {\bf a} \cdot {\bf \nabla}_u \aleph_a \, d \hat{\Omega}_u
& = & \int_{\hat{\Omega}_u} {\bf \nabla}_u \cdot ( {\bf a} \aleph_a ) \, d \hat{\Omega}_u
\; - \; \int_{\hat{\Omega}_u} ( {\bf \nabla}_u \cdot {\bf a} ) \aleph_a \, d \hat{\Omega}_u
\end{eqnarray}
Gauss' law then can be used to show that the first term in
(\ref{int_by_parts}) vanishes because $\aleph_a$ vanishes on the boundary
$\partial \hat{\Omega}_u$
\begin{eqnarray*}
\int_{\hat{\Omega}_u} {\bf \nabla}_u \cdot ( {\bf a} \aleph_a ) \, d \hat{\Omega}_u
& = & \oint_{\partial \hat{\Omega}_u} \aleph_a {\bf a} \cdot d{\bm \Sigma}_u
\; = \; 0
\end{eqnarray*}
where $d{\bm \Sigma}_u$ is the three-volume element on $\partial
\hat{\Omega}_u$.  The kernel ${\bf \nabla}_u \cdot {\bf a}$ in the second term
in (\ref{int_by_parts}) also vanishes because the Faraday tensor is
antisymmetric and independent of the velocity coordinate\footnote{In
component notation,
\begin{eqnarray*}
\frac{\partial}{\partial {\rm u}^{\mu}} \left( {\rm u}^{\nu} {F^{\mu}}_{\nu} 
\right) & = & {\delta^{\nu}}_{\mu} \, {F^{\mu}}_{\nu} 
\; = \; {F^{\mu}}_{\mu} \; = \; 0
\end{eqnarray*}}
\begin{eqnarray*}
{\bf \nabla}_u \cdot \left( \frac{q_a {\bf u} \cdot {\bf F}}{m_a c} \right) & = 
& 0
\end{eqnarray*}
Similarly the velocity integral of the right-hand side of equation
(\ref{geom_rel_boltz}) vanishes because $\dot{\aleph}_{a, col}$ is in a
similar form to the ${\bf a} \cdot {\bf \nabla}_u \aleph_a$ term in equation
(\ref{geom_rel_boltz}).  (See equation \ref{dens_coll}.)  Because ${\bf
x}$ and ${\bf u}$ are independent generalized coordinates, ${\bf \nabla}_x
\cdot {\bf u} = 0$, so the velocity integral of the {\em zeroth} moment
of equation (\ref{geom_rel_boltz}) becomes simply \begin{eqnarray}
\label{raw_plasma_cont}
{\bf \nabla}_x \cdot \int_{\hat{\Omega}_u} {\bf u} \aleph_a \, d\hat{\Omega}_u & = & 0
\end{eqnarray}
In order to interpret this equation, we need a coordinate gauge in which
to express ${\bf u}$.

\subsection{Velocity Decomposition and the Velocity Coordinate Gauge}

We choose to decompose the velocity coordinate ${\bf u}$ into four
components, one along the center-of-rest-mass average particle velocity
${\bf{U}}$
\begin{eqnarray}
\label{avg_vel_def}
{\bf{U}} & \equiv & 
\frac{\sum_a m_a \int_{\hat{\Omega}_u} {\bf u} \aleph_a \, d\hat{\Omega}_u}
{\sum_a m_a \int_{\hat{\Omega}_u} \aleph_a \, d\hat{\Omega}_u}
\end{eqnarray}
(the justification for this choice of average velocity will be given
later) and one orthogonal to ${\bf{U}}$ (the drift four-velocity ${\bf v}$), 
yielding 
\begin{eqnarray}
\label{vel_decomp}
{\bf u} & = & \gamma \, ( {\bf{U}} + {\bf v} )
\end{eqnarray}
where the Lorentz factor is
\begin{eqnarray}
\label{lorentz_factor}
\gamma & \equiv & - \frac{1}{c^2} ( {\bf{U}} \cdot {\bf u} )
\end{eqnarray}
and the relative spatial velocity coordinate is 
\begin{eqnarray}
{\bf v} & \equiv & ( {\bf{P}} \cdot {\bf u} ) / \gamma
\end{eqnarray}
Note that ${\bf v}$ is still a four-vector, but it is constrained to have
only three independent components by its orthogonality with ${\bf{U}}$
\begin{eqnarray}
{\bf{U}} \cdot {\bf v} & = & 0
\end{eqnarray}
It therefore is a velocity {\em coordinate} that spans $\Upsilon_u$. 
In the rest frame of the fluid, the velocity components are
\begin{eqnarray}
{\bf u} & = & \left( \gamma \, c, \; \gamma {\rm v}^1, \; \gamma {\rm v}^2, \; 
\gamma {\rm v}^3 \right)
\\
{\bf v} & = & \left( 0, \; {\rm v}^1, \; {\rm v}^2, \; {\rm v}^3 \right)
\end{eqnarray}
and the constraint (equation \ref{mass_hyperboloid}) on the particle
velocity ${\bf u}$ becomes
\begin{eqnarray*}
\gamma & = & \left( 1 - {\bf v} \cdot {\bf v} \right)^{-1/2}
\end{eqnarray*}
as expected.  

With the velocity decomposition in equation (\ref{vel_decomp}), the
equation of continuity (\ref{raw_plasma_cont}) for particle species
'$a$' becomes
\begin{eqnarray}
\label{plasma_cont}
{\bf \nabla}_x \cdot n_a ( {\bf{U}} + {\bf{V}}_a ) & = & 0
\end{eqnarray}
where
\begin{eqnarray}
\label{spec_dens}
n_a & \equiv & \int_{\hat{\Omega}_u} \gamma \aleph_a \, d\hat{\Omega}_u
\end{eqnarray}
is the particle density and 
\begin{eqnarray}
\label{spec_vel}
{\bf{V}}_a & = & \frac{1}{n_a} \, \int_{\hat{\Omega}_u} \gamma
\, {\bf v} \aleph_a \, d\hat{\Omega}_u
\end{eqnarray}
is the average particle {\em drift} velocity for species '$a$'.
The equation of continuity (\ref{plasma_cont}) is important for the
conservation laws of rest mass and charge in one-fluid dynamics.  Despite
the Lorentz factor in the above velocity integrals, these quantities are,
in fact, the familiar three-momentum integrals of standard thermodynamics,
in which the single factor of $\gamma$ does not appear. (See Appendix
\ref{three_d_integrals}).

\subsection{The First Moment:  Conservation of Particle Energy-Momentum}

The first moment of equation (\ref{geom_rel_boltz}) generates a vector
equation
\begin{eqnarray}
\label{first_moment}
{\bf \nabla}_x \cdot ( \aleph_a \, {\bf u} \otimes {\bf u} ) \; + \; 
{\bf u} \left\{ \frac{q_a \, {\bf u} \cdot {\bf F}}{m_a c} \, \cdot \, 
{\bf \nabla}_u \aleph_a \right\} & = & {\bf u} \, \dot{\aleph}_{a, col}
\end{eqnarray}
which also can be integrated over $d\hat{\Omega}_u$ to yield
\begin{eqnarray}
\label{plasma_energy_mom}
{\bf \nabla}_x \cdot \left[ {n'}_a \, {\bf{U}} \otimes {\bf{U}}  \; + \; 
n_a \, {\bf{U}} \otimes {{\bf{V}}'}_a \; + \; 
n_a \, {{\bf{V}}'}_a \otimes {\bf{U}} \; + \; {\Pi}_a \right]
& = & \frac{1}{m_a c} {\bf{J}}_a \cdot{\bf F} \; - \; 
\nu \, n_a ( {\bf{U}} + {\bf{V}}_a )
\end{eqnarray}
where
\begin{eqnarray} 
\label{enh_spec_dens}
{n'}_a & \equiv & \int_{\hat{\Omega}_u} \gamma^2 \aleph_a \, d\hat{\Omega}_u
\\
\label{enh_spec_vel}
{{\bf{V}}'}_a  & \equiv & \frac{1}{n_a} 
\int_{\hat{\Omega}_u} \gamma^2 \, {\bf v} \aleph_a \, d\hat{\Omega}_u
\end{eqnarray}
In deriving equation (\ref{plasma_energy_mom}) we have substituted
equation (\ref{vel_decomp}) for ${\bf u}$ in the first term of
(\ref{first_moment}) and discarded a vanishing boundary integral that
results from integrating the second term by parts.

Note the extra factor of $\gamma$ in the integrals in equations
(\ref{enh_spec_dens}) and (\ref{enh_spec_vel}) compared with
(\ref{spec_dens}) and (\ref{spec_vel}).  These are beamed quantities
that give rise to relativistic internal energy, pressure, {\it etc.}
The partial current is
\begin{eqnarray}
{\bf{J}}_a & \equiv & q_a \, \int_{\hat{\Omega}_u} {\bf u} \aleph_a \, d\Omega 
\; = \; q_a \, n_a \, ( {\bf{U}} + {\bf{V}}_a )
\end{eqnarray}
which has components $q_a \, n_a \, ( c, {{\rm V}^1_a}, {{\rm V}^2_a}, {{\rm V}^3_a} )$
in the fluid rest frame.  A four-vector
\begin{eqnarray}
{\bf j}_a & = & q_a \, n_a \, {\bf{V}}_a
\end{eqnarray}
can be used to describe the ${\bf U}$-orthogonal  part of the current ({\it cf.},
equation \ref{spatial_current}), with ${\bf{U}} \cdot {\bf j}_a = 0$.
The partial pressure tensor (per unit mass) of species '$a$' is
\begin{eqnarray}
{\Pi}_a & \equiv & \int_{\hat{\Omega}_u} ( {\bf{P}} \cdot {\bf v} ) \otimes 
( {\bf{P}} \cdot {\bf v} ) \, \aleph_a \, d\hat{\Omega}_u
\nonumber
\\
& = & \int_{\hat{\Omega}_u} \gamma^2 \, ( {\bf v} \otimes {\bf v} ) \, 
\aleph_a \, d\hat{\Omega}_u
\end{eqnarray}
Note also in equation (\ref{plasma_energy_mom}) that the collision term
has been simplified to a collision frequency $\nu$ times the integrated
particle flux.

\section{The One-Fluid Equations}
\label{onefl}

At present, astrophysical simulation codes deal almost exclusively with the one-fluid
equations when dynamics are concerned.  That is, the individual particle
species dynamical equations are summed and solved as a single set of
equations.  Of course, many stellar evolution and collapse codes track the
composition for different species of particle, but this is done usually
for equation of state and composition purposes only, not to determine
relative drift velocities of different charged species, for example.
Therefore, for at least the near future, it still will be important to 
construct one-fluid equations for studies of processes like black hole 
formation and accretion.

\subsection{The Example of Hydrodynamics}

Before deriving the equations of relativistic charge dynamics, it is
important to review the derivation of the equations of hydrodynamics
in the presence of an electromagnetic field.  This is more than just
an exercise in ``re-inventing the wheel''.  Reviewing this derivation
will allow us to check that our equations and procedures are correct and 
define some quantities that will be needed later in the discussion of
charge dynamics, and it will assist us in understanding the new set of
equations in terms of the familiar hydrodynamic ones.  The procedure
is to simply weight the five equations (\ref{plasma_cont}) and
(\ref{plasma_energy_mom}) by the particle {\em rest} mass $m_a$ and sum
over all species
\begin{eqnarray}
\label{hyd_cont}
{\bf \nabla}_x \cdot \rho_m {\bf{U}} & = & 0
\\
\label{hyd_energy_mom}
{\bf \nabla}_x \cdot \left\{ [ \rho_m \, + \, \varepsilon/c^2 ] 
{\bf{U}} \otimes {\bf{U}} \; + \; \frac{1}{c^2} \left[ {\bf{U}} \otimes {\bf{H}} 
\; + \; {\bf{H}} \otimes {\bf{U}} \right] \; + \; p {\bf{P}} \right\} & = & 
\frac{1}{c} {\bf{J}} \cdot {\bf F}
\end{eqnarray}
where
\begin{eqnarray}
\rho_m & \equiv & \sum_a n_a \, m_a
\end{eqnarray}
is the total {\em rest} mass density, 
\begin{eqnarray}
\label{hyd_energy}
\varepsilon & \equiv & \sum_a \varepsilon_a \; \equiv \; 
\sum_a ( {n'}_a - n_a ) \, m_a c^2 \; = \; 
\sum_a m_a c^2 \int_{\hat{\Omega}_u} \gamma \, ( \gamma - 1 ) \aleph_a \, d\hat{\Omega}_u
\end{eqnarray}
is the internal (kinetic) energy density, 
\begin{eqnarray}
{\bf{H}} & \equiv & \sum_a n_a \, m_a c^2 \, {{\bf{V}}'}_a  
\end{eqnarray}
is the heat flux (including relativistic enhancement), and 
\begin{eqnarray}
\label{hyd_pressure}
p & \equiv & \sum_a p_a \; \equiv \; \sum_a \, \frac{m_a}{3} 
\int_{\hat{\Omega}_u} \gamma^2 ( {\bf v} \cdot {\bf v} ) \aleph_a d\hat{\Omega}_u
\end{eqnarray}
is the scalar pressure for an {\em isotropic} distribution in
$\Upsilon_u$.  Equation (\ref{hyd_cont}) is correct only if the
rest-mass-centered drift velocity is zero
\begin{eqnarray*}
\sum_a n_a \, m_a \, {\bf{V}}_a & = & 0
\end{eqnarray*}
and that is the case only if the average velocity is defined according to
equation (\ref{avg_vel_def}), thereby justifying our rest-mass-centered
choice for ${\bf{U}}$.

We can show that equation (\ref{hyd_energy_mom}) is equivalent to equation
(\ref{divTzero}) if we define the following energy-momentum-stress
tensor\footnote{In component notation the inner and scalar products of
two tensors are
\begin{eqnarray*}
[ {\bf F} \cdot {\bf G} ]^{\alpha \beta} & = & F^{\alpha \nu} 
{G^{\beta}}_{\nu}
\\
{\bf F} {\bf :} {\bf G} & = & F^{\mu \nu} \, G_{\mu \nu}
\end{eqnarray*}}
\begin{eqnarray} 
{\bf{T}} & \equiv & {\bf{T}}^{FL} \; + \; {\bf{T}}^{EM}
\\
\label{T_fl}
{\bf{T}}^{FL} & \equiv & [ \rho_m + ( \varepsilon + p ) / c^2 ] \, 
{\bf{U}} \otimes {\bf{U}} \; + \; \frac{1}{c^2} \left[ {\bf{U}} \otimes {\bf{H}} \; + \; 
{\bf{H}} \otimes {\bf{U}} \right] \; + \; p \, {\bf g} 
\\
{\bf{T}}^{EM} & \equiv & \frac{1}{4 \pi} \left[ {\bf F} \cdot {\bf F} 
\; - \; \frac{1}{4} \, ({\bf F} {\bf :} {\bf F} ) {\bf g} \right]
\end{eqnarray}
and recognize from equation (\ref{faraday}) that
\begin{eqnarray}
\frac{1}{c} {\bf{J}} \cdot {\bf F} & = & - {\bf \nabla}_x \cdot {\bf{T}}^{EM}
\end{eqnarray}
The multi-fluid equations of section \ref{multifl}, therefore, reproduce
the familiar equations of general relativistic hydrodynamics.

Note that the energy-momentum-stress tensor in equation (\ref{T_fl}) is
the one for an ideal gas {\em with} heat flow.  It does {\em not} contain
terms for viscous momentum and energy transport, however.  These latter 
terms would arise if we performed a more sophisticated treatment of the 
collision term.  
Also, unlike nonrelativistic
treatments (which require the second moment for the energy equation),
the treatment here derives the conservation of energy using only the
first moment.  This occurs because the equations are relativistic 
and use the four-vector ${\bf u}$ in the first moment instead of the
three-velocity ${\bf v}$.  The conservation of energy equation can be
extracted by taking the component of equation (\ref{hyd_energy_mom})
along the average velocity\footnote{The Euler equations also can be extracted
by projecting equation (\ref{hyd_energy_mom}) with the projection tensor
${\bf{P}}$, but they are of no interest in this paper.}
\begin{eqnarray}
\label{hyd_first_law}
{\bf{U}} \cdot {\bf \nabla}_x \varepsilon \; + \; 
( \varepsilon + p ) {\bf \nabla}_x \cdot {\bf{U}} & = & 
- {\bf j} \cdot {\bf F} \cdot {\bf{U}} \; - \; {\bf \nabla}_x \cdot {\bf{H}}
\; - \; {\bf{H}} \cdot ( {\bf{U}} \cdot {\bf \nabla}_x {\bf{U}} ) / c^2
\end{eqnarray}
which is also known as the ``first law of thermodynamics'':  the change
in internal energy $\varepsilon$ plus mechanical work is given by Ohmic
heating minus losses due to heat conduction.

\subsection{Charge Dynamics in Geometric form}

A similar set of equations to (\ref{hyd_cont}) and (\ref{hyd_energy_mom})
can be generated by weighting (\ref{plasma_cont}) and
(\ref{plasma_energy_mom}) with the particle charge $q_a$ rather than
the rest mass.  The results are the equations of charge dynamics:  the
conservation of charge
\begin{eqnarray}
\label{chg_cont}
{\bf \nabla}_x \cdot {\bf{J}} & = & {\bf \nabla}_x \cdot ( \rho_q {\bf{U}} + 
{\bf j} ) \; = \; 0
\end{eqnarray}
(equivalent to equation \ref{divJzero}) and the relativistic generalized Ohm's law
\begin{eqnarray}
\label{chg_energy_mom}
{\bf \nabla}_x \cdot {\bf{C}} & = & \ell \left[ \frac{1}{c} \left( \bf{U} \, + \, 
{\it h} \, {\bf j} \right) \cdot {\bf F} 
\; - \; \eta \, ( \rho_q {\bf{U}} + {\bf j} ) \right]
\end{eqnarray}
where the charge-current-pressure tensor is given by
\begin{eqnarray}
\label{chg_curr_tensor}
{\bf{C}} & \equiv & [ \rho_q \, + \, ( \varepsilon_q + p_q ) / c^2 ] \,  
{\bf{U}} \otimes {\bf{U}} \; + \; {\bf{U}} \otimes {\bf j}' 
\; + \; {\bf j}' \otimes {\bf{U}} \; + \; p_q {\bf g}
\end{eqnarray}
{\em Note the appearance of a beamed current ${\bf j}'$ in the
charge-current-pressure tensor while the source terms involve the unbeamed
current ${\bf j}$ only.}

The individual charge-dynamical scalars are charge density,
charge-weighted internal energy and pressure (per unit mass),
\begin{eqnarray}
\label{chg_dens}
\rho_q & \equiv & \sum_a n_a \, q_a
\\
\label{chg_energy}
\varepsilon_q & \equiv & \sum_a \frac{q_a}{m_a} \varepsilon_a \; = \; 
\sum_a ( {n'}_a - n_a ) \, q_a c^2 \; = \; 
\sum_a q_a c^2 \int_{\hat{\Omega}_u} \gamma \, ( \gamma - 1 ) \aleph_a \, d\hat{\Omega}_u
\\
\label{chg_pressure}
p_q & \equiv & \sum_a \frac{q_a}{m_a} p_a \; = \; \frac{1}{3} \, \sum_a q_a 
\int_{\hat{\Omega}_u} \gamma^2 ( {\bf v} \cdot {\bf v} ) \aleph_a \, d\hat{\Omega}_u
\end{eqnarray}
and an enhancement $\gamma_q$ in the spatial electric current due to 
relativistic beaming effects
\begin{eqnarray}
\label{chg_beaming}
{\bf j}' & \equiv & 
\sum_a q_a \int_{\hat{\Omega}_u} \gamma^2 \, {\bf v} \aleph_a \, d\hat{\Omega}_u
\; \equiv \;
\sum_a q_a \, n_a \, {{\bf{V}}'}_a 
\; \equiv \; 
\gamma_q \, \sum_a q_a \, n_a \, {\bf{V}}_a 
\; = \; 
\gamma_q \, {\bf j} 
\end{eqnarray}
Because the partial internal energies $\varepsilon_a$ and pressures
$p_a$ have been defined previously (equations \ref{hyd_energy}
and \ref{hyd_pressure}), and because $q_a$ and $m_a$ are known, the
quantities $\varepsilon_q$ and $p_q$ (equations \ref{chg_energy} and
\ref{chg_pressure}) are not new variables but rather different weightings
of known equations of state.  Only the six quantities $\rho_q$, ${\bf
j}'$, and $\gamma_q$ are new ones that need their own charge-dynamical
equations. Five of those equations are, respectively, (\ref{chg_cont}),
the three components of (\ref{chg_energy_mom}) orthogonal to ${\bf{U}}$,
and the component of (\ref{chg_energy_mom}) parallel to ${\bf{U}}$.
Because ${\bf j}'$ is a four-vector, the sixth equation is a constraint
on its components
\begin{eqnarray}
\label{jprime_orthog}
{\bf U} \cdot {\bf j}' & = & 0
\end{eqnarray}
similar to equation (\ref{j_orthog}).

The sources and sinks of current on the right-hand side of equation
(\ref{chg_energy_mom}) are the Lorentz effect, the Hall effect, and
resistive losses, with coefficients
\begin{eqnarray}
\ell & \equiv & \sum_a ( q_a^2 \, n_a / m_a )
\\
{\it h} & \equiv & \frac{1}{ \ell \, {|{\bf j}|}} \sum_a \frac{q_a}{m_a} \, 
| {\bf j}_a | ) 
\\
\eta & \equiv & \frac{\nu}{\ell}
\end{eqnarray}
where ${|{\bf j}|} \equiv (- {\bf j} \cdot {\bf j} )^{1/2}$ is the
magnitude of the spatial current.

Note that the definition of $\gamma_q$ (equation \ref{chg_beaming}) makes
use of the fact that ${{\bf{V}}'}_a$ and ${\bf{V}}_a$ are essentially
parallel, resulting in the current ${\bf j}$ being enhanced by an
average Lorentz factor $\gamma_q$ (which is $\ge 1$).  We can derive an
equation for $\gamma_q$ in a manner similar to that used for equation
(\ref{hyd_first_law}), arriving at
\begin{eqnarray}
{\bf j}' \cdot {\bf \nabla}_x ( \gamma_q ) & = & 
\gamma_q^2  \; \left\{ \; \ell \, \eta \, \rho_q 
\; - \; \frac{1}{c^2} \left[ {\bf j}' \cdot ( {\bf{U}} \cdot {\bf \nabla}_x {\bf{U}} ) 
\; + \; {\bf{U}} \cdot {\bf \nabla}_x \varepsilon_q 
\; + \; ( \varepsilon_q + p_q ) {\bf \nabla}_x \cdot {\bf{U}} \right] 
\; \right\} 
\nonumber
\\
\label{chg_first_law}
& & 
\; - \; \gamma_q ( \gamma_q - 1 ) \, {\bf \nabla}_x \cdot {\bf j}' 
\; - \; \frac{\gamma_q {\it h} \ell}{c^3} \, {\bf j}' \cdot {\bf F} \cdot {\bf{U}} 
\end{eqnarray}
% with $\gamma_q$ as the only unknown.  
While analogous to equation
(\ref{hyd_first_law}), equation (\ref{chg_first_law}) is of a very
different character.  Everything in it, including ${\bf j}'$, $\rho_q$,
${\bf U}$, ${\bf F}$, and even $\varepsilon_q$ and $p_q$ can be considered
known.  What remains is a current-weighted gradient of the Lorentz factor
equaling a quadratic function of that Lorentz factor. In the rest frame
of the fluid (or when the fluid is at rest), the gradient ${\bf j}'
\cdot {\bf \nabla}_x$ loses all time dependence, and {\em the equation takes
on an elliptical character}.  The distribution of $\gamma_q$ must be
solved implicitly on each hypersurface, employing appropriate boundary
conditions, {\it etc}.  Equation (\ref{chg_first_law}) can be thought
of as a constraint on the current beaming factor $\gamma_q$, enforcing
the conservation of charge-weighted internal energy flow via ${\bf j}'$
at the same time as conservation of charge flow via ${\bf j}$ is enforced
by equation (\ref{chg_cont}).

Implicit equations for Lorentz factors are not unusual in relativistic
hydrodynamics or magneto-hydrodynamics \cite{dh94, mm99, hmd02, k03}.
In most cases, however, they are simple algebraic equations that need
to be solved in a single cell in spacetime at each time step.  In this
case, the equation contains gradient information on $\gamma_q$ and,
therefore, must be solved over the entire hypersurface simultaneously.
When the fluid is not at rest with respect to the frame in which the
gradient ${\bf \nabla}_x$ is computed, and is flowing relativistically,
${\bf j}'$ can have a significant temporal component, rendering
equation (\ref{chg_first_law}) an hyperbolic equation.  Nevertheless,
computationally, it would be wise to solve this particular equation
implicitly at all time steps in order to avoid numerical problems when
the local fluid velocity slows down.

\subsection{Charge Dynamics in Component Form}

The equations of charge dynamics (\ref{chg_cont}-\ref{chg_curr_tensor}) 
are valid in any frame.  Therefore, we can immediately convert them to
component form in any metric.  Using $\rho_q$, ${\bf j}'$, and $\gamma_q$
as the variables, they are
\begin{eqnarray}
\label{comp_chg_cont}
\left[ ( \, \rho_q \, {\rm U}^{\mu} \; + \; {{\rm j}'}^{\mu} / \gamma_q \, ) 
\, \sqrt{-g} \right]_{, \; \mu} & = & 0
\\
\label{comp_chg_energy_mom}
( \, {\rm C}^{\alpha \mu} \, \sqrt{-g} \, )_{, \; \mu} 
\; + \; {\Gamma^{\alpha}}_{\mu \nu} \, {\rm C}^{\mu \nu} \, \sqrt{-g} 
& = & \ell \, \left[ \frac{1}{c} \left( \, {\rm U}^{\mu} \, + \, \frac{h \, {{\rm j}'}^{\mu}}{\gamma_q} \, \right)
{{\rm F}^{\alpha}}_{\mu} 
\; - \; \eta \, \left( \, \rho_q \, {\rm U}^{\alpha} \, + \, \frac{{{\rm j}'}^{\mu}}{\gamma_q} \, \right)
\right] \, \sqrt{-g} ~~~~~~~~~~
\\
{{\rm j}'}^{\mu} \, {\rm U}^{\nu} \, g_{\mu \nu} & = & 0
\end{eqnarray}
with the charge-current-pressure tensor given by
\begin{eqnarray}
\label{comp_chg_curr_tensor}
{\rm C}^{\alpha \beta} & = &
\left[ \, \rho_q \, + \, ( \varepsilon_q + p_q ) / c^2 \right] 
\, {\rm U}^{\alpha} \, {\rm U}^{\beta} 
\; + \; {\rm U}^{\alpha} \, {{\rm j}'}^{\beta} 
\; + \; {{\rm j}'}^{\alpha} \, {\rm U}^{\beta} \; + \; p_q \, g^{\alpha \beta}
\end{eqnarray}
and the usual Einstein summation convention and comma derivative applying
\begin{eqnarray*}
{{\rm j}'}^{\mu} \, {\gamma_q}_{, \; \mu}
& \equiv & \sum_{\mu = 0}^3 \, {{\rm j}'}^{\mu} \, \frac{\partial \gamma_q}{\partial x^{\mu}}
\end{eqnarray*}

In addition, we can replace the zeroth (temporal) component of equation 
(\ref{comp_chg_energy_mom}) with that projected along the four-velocity to get 
a component form of equation (\ref{chg_first_law}) for the current beaming factor
\begin{eqnarray}
{{\rm j}'}^{\mu} \, {\gamma_q}_{, \; \mu} 
& = & \gamma_q^2 \, \left\{ \, \ell \, \eta \, \rho_q \right.
\nonumber
\\
& & - \; \frac{1}{c^2} \left[ {{\rm j}'}^{\mu} \, g_{\mu \nu} \, {\rm U}^{\lambda} 
\, ( \, {{\rm U}^{\nu}}_{, \; \lambda} \, + \, {\Gamma^{\nu}}_{\sigma \lambda} 
\, {\rm U}^{\sigma} \, ) 
\; + \; {\rm U}^{\mu} \, {{\varepsilon}_q}_{, \; \mu}
\; + \; \frac{(\varepsilon_q + p_q )}{\sqrt{-g}} 
\, ( \, {\rm U}^{\mu} \, \sqrt{-g} \, )_{, \; \mu} \right]
\nonumber
\\
\label{comp_chg_first_law}
& & - \left. \; \frac{\gamma_q(\gamma_q-1)}{\sqrt{-g}} ( \, {{\rm j}'}^{\mu} \, \sqrt{-g} \, )_{, \; \mu}
\; - \; \frac{\gamma_q \, h \, \ell}{c^3} \, {{\rm j}'}^{\mu} \, F_{\mu \nu} \, {\rm U}^{\nu}
\right\}
\end{eqnarray}

\section{Discussion}
\label{discussion}

\subsection{Special Cases} It is useful to examine a few special cases
of relativistic charge dynamics to compare with previous authors' work.

\subsubsection{Steady State with No Hall Term}

Under many conditions the terms on the left hand side of
(\ref{chg_energy_mom}) are small compared to those on the right hand
side, because the time and length scales over which plasma properties
vary are long.  In addition, the Hall effect is often small compared
to the Lorentz and resistive effects.  The remaining terms, when 
projected orthogonal to ${\bf U}$, then reduce to equation (\ref{relmhd}).   
They can be reduced further to the time-independent relativistic and 
nonrelativistic forms in section \ref{problems} under conditions of 
infinite conductivity, sub-relativistic flow, {\it etc}.

\subsubsection{Cold Plasma and Nonrelativistic Flow}

Equation (\ref{chg_energy_mom}) also can be reduced to the form
(\ref{relohmslaw}) if we make the following cold plasma assumptions: 
1) the charge-weighted internal energy and pressure are negligible
compared with $\rho_q c^2$ and 
2) the current beaming factor $\gamma_q = 1$ so that ${\bf j}' = {\bf j}$. 
These conditions allow the charge-current-pressure tensor to be
re-written as
\begin{eqnarray*}
{\bf{C}} & \equiv & 
\;  {\bf{U}} \otimes {\bf J} \; + \; {\bf J} \otimes {\bf{U}} 
\; - \; \rho_q \, {\bf{U}} \otimes {\bf{U}} 
\; + \; p_q {\bf g}
\end{eqnarray*}
and allow us to ignore the temporal component of (\ref{chg_energy_mom}),
which is now redundant with the conservation of charge equation. 
With the assumption that the metric is that of Minkowskian flat space 
(with non-zero components $g^{\alpha \alpha} = [-1,~1,~1,~1]$, no sum 
on $\alpha$), the three spatial components of (\ref{chg_energy_mom}) 
(those projected orthogonal to ${\bf e}_t$, {\em not} ${\bf e}_{\bf U}$) 
to reduce to equation (\ref{relohmslaw}).

If we additionally make the following nonrelativistic assumptions that 
${\bf U} = [c,~V_x,~V_y,~V_z]$ and $|{\bm V}| << c$, so that the
orthogonal current {\bf j} is approximately the spatial current ${\bm J} =
[0,~J_x,~J_y,~J_z]$, then equation (\ref{chg_energy_mom}) then reduces 
to equation (\ref{nonrelohmslaw}).

\subsubsection{Relativistic Pair Plasma}

For highly relativistic flows a time-dependent form of equation
(\ref{chg_energy_mom}) will be needed.  However, for certain plasmas,
such as a relativistic pair plasma near a black hole, equation
(\ref{chg_first_law}) still can be simplified somewhat.  In this
case, $m_{-} = m_{+}$ and $q_{-} = - q_{+}$, so $\varepsilon_q =
p_q = 0$, and the Hall coefficient vanishes explicitly. Then equation
(\ref{chg_first_law}) reduces to
\begin{eqnarray}
\label{pair_plasma}
{\bf j}' \cdot {\bf \nabla}_x ( \gamma_q ) & = & 
\gamma_q^2  \; [ \; \ell \, \eta \, \rho_q 
\; - \; {\bf j}' \cdot ( {\bf{U}} \cdot {\bf \nabla}_x {\bf{U}} ) / c^2
\; ] 
\; - \; \gamma_q ( \gamma_q - 1 ) {\bf \nabla}_x \cdot {\bf j}' 
\end{eqnarray}
In other words, the beaming factor distribution is determined by the
competition between local `creation' of charge by collisions and the
loss of charge through the beamed current ${\bf j}'$.

\subsubsection{Uniform Adiabatic Index and Mixture}

More generally, if the plasma is made up of particle partial fluids
that have the same adiabatic index, and if that index and the fractional
pressure $\pi_a \equiv p_a / p$ of each species are uniform throughout $\Omega_x$, then
we have each quantity $\varepsilon_a$, $\varepsilon$, and $\varepsilon_q$
related to their respective pressures by the simple relation
\begin{eqnarray}
\varepsilon_i & = & \frac{1}{\Gamma - 1} \, p_i
\end{eqnarray}
where $i = a$, $q$, or blank and $\Gamma$ is the adiabatic index.
Then the ratio of charge-weighted to total pressure (and that for internal
energy) is a uniform constant throughout $\Omega_x$
\begin{eqnarray}
\frac{p_q}{p} & = & \frac{\varepsilon_q}{\varepsilon} 
\; = \; \sum_a \, \frac{q_a}{m_a} \, \pi_a \; \equiv \; \zeta
\end{eqnarray}
We then can use equation (\ref{hyd_first_law}) to eliminate the
charge-weighted thermodynamic quantities in (\ref{chg_first_law}) to
obtain a simpler equation for $\gamma_q$
\begin{eqnarray}
{\bf j}' \cdot {\bf \nabla}_x ( \gamma_q ) & = & 
\gamma_q^2  \; [ \; \ell \, \eta \, \rho_q 
\; - \; ({\bf j}' - \zeta {\bf{H}}/c^2 ) \cdot ( {\bf{U}} \cdot {\bf \nabla}_x {\bf{U}} ) / c^2
\; - \; \, {\bf \nabla}_x \cdot ( {\bf j}' - \zeta {\bf{H}}/c^2 )
\; ] 
\nonumber
\\
& & 
\label{unif_adiab}
\; + \; \gamma_q \; [ \; {\bf \nabla}_x \cdot {\bf j}' 
\; - \; \frac{( {\it h} \ell - \zeta )}{c^3} \, {\bf j}' \cdot {\bf F} \cdot {\bf{U}} \; ]
\end{eqnarray}
The pair plasma is a special case of these conditions, and equation
(\ref{unif_adiab}) reduces to (\ref{pair_plasma}) in this case ($\zeta$,
$h$ $\rightarrow 0$).  Equation (\ref{unif_adiab}) is useful for showing
how the equation for $\gamma_q$ is decoupled from the terms involving
$\varepsilon_q$ in equation (\ref{chg_first_law}).  The quantity ${\bf
j}' - \zeta {\bf H}/c^2$ is a residual (beamed) current whose properties
can affect the value of the current beaming factor $\gamma_q$.

\subsection{One-fluid {\it vs.} Multi-fluid Theory}

The advantage of one-fluid theory, of course, is that, by summing the
multi-fluid equations and deriving thermodynamic quantities that close
the system, the many equations for each particle species are reduced to
only five (\ref{hyd_cont} and \ref{hyd_energy_mom}).
%, and \ref{hyd_first_law}).
However, with the introduction of a set of five new one-fluid equations
(\ref{chg_cont} and \ref{chg_energy_mom}), one
%, and \ref{chg_first_law}), one
must ask whether it is still useful to deal with the one-fluid equations
rather than the more instructive multi-fluid equations, particularly if
there are only two fluids (ions and electrons).

The answer still appears to be a qualified `yes', although it is quite 
likely that the multi-fluid equations will become more important in the 
next few decades, if not sooner.  First of all, most current astrophysical 
codes (hydrodynamic and magnetohydrodynamic, mainly) are one-fluid codes.
While nevertheless a significant task, the addition of the one-fluid
charge dynamical equations to existing MHD codes is still much less
effort than developing an entirely new multi-fluid code.  Secondly,
even when constructing a new code that involves only ions and electrons,
developing that two-fluid code will be a much greater task than developing
a one-fluid code with the ten equations of hydro- and charge dynamics.
In one-fluid theory, the collision terms can be treated with much less
rigor than in a two-fluid code.  In the former, because thermodynamic
equilibrium is assumed, one need only postulate an approximate
resistivity, as we have done in section \ref{onefl}.  In the latter
case, however, one must carefully handle collisional momentum and energy
transfer between each species, as well as the scattering of particles by
electromagnetic oscillations with wavelengths shorter than a cell size.
Otherwise, the simulated multi-temperature structure of the fluid will
be meaningless.  Finally, as is the case in simulations of late stages
of stellar evolution and collapse, when dealing with the collapse of dense
matter to black holes, there probably will be many more than two species
of particle (neutrons, protons, electrons, positrons, heavy iron-peak
nuclei, {\it etc.}).  With five multi-fluid equations for each species,
the number of equations to integrate could be significantly greater than
the ten needed for hydro- and charge dynamics.

Therefore, any multi-fluid astrophysical codes that are to be developed
in the next few years are likely to be two-fluid only and probably will
make the assumption of thermodynamic equilibrium initially.  In that case
their results will be similar to those obtained by older MHD codes that
have been modified to handle charge dynamics. Nevertheless, these new
codes will become increasingly sophisticated as more particle physics is
added and should lead to a greater understanding of black hole formation
than is possible from one-fluid simulations alone.

\section{Conclusions}

This paper has used geometric frame-independent tensor notation to
derive what the author believes is the first set of one-fluid equations 
to be published that correctly describes the flow of charge in general 
relativistic environments.  
Previously used or suggested approximations ({\it e.g.}, 
force-free field, magnetohydrodynamics, even currently-available relativistic 
forms of the generalized Ohm's law) do not correctly describe black hole 
astrophysics on time scales much shorter than the collapse time ($<< GM/c^3$) 
or in strong gravitational fields.  
The principle equations of charge dynamics are (\ref{chg_cont}) and 
(\ref{chg_energy_mom}), with an alternative form for the relativistic current 
equation (\ref{chg_first_law}). Proper handling of charge flow in such
environments will be important for understanding the details of highly 
relativistic astrophysical events like black hole formation and relativistic 
jet generation, which may be important for understanding gamma-ray bursts, 
{\it etc.}  (Multi-fluid equations are
also derived, but the collision terms are not treated with sufficient
rigor in this paper to make them useful for detailed simulations
at this time.)  These equations of `charge dynamics', also given in
component form in (\ref{comp_chg_cont}, \ref{comp_chg_energy_mom}, and
\ref{comp_chg_first_law}), are valid for any flow velocity (causal),
in any reference frame (covariant), and in any spacetime metric
(general relativistic). They therefore are suitable not only for flow
in stationary metrics like Schwarzschild or Kerr, but also for general
numerical relativity calculations that include a fluid with an embedded
electromagnetic field.  The equations of charge dynamics were shown to
reduce to a variety of prior ``generalized Ohm's laws'' in the appropriate
cold-plasma, flat-space or nonrelativistic limits.

In addition to the general relativistic nature of the equations, the
principal difference between this paper and previous work is that it does
{\em not} make the assumption of a cold plasma.  Not only must one deal 
with quantities such as charge-weighted internal energy and pressure,
one also must solve for the current beaming factor $\gamma_q$ that 
distinguishes the beamed current ${\bf j}'$ from the unbeamed ${\bf j}$. 
The equation for $\gamma_q$, generated by the subtraction of the energy
equations for positive and negative ions, is primarily implicit and
global in character, providing a constraint on the flow of charge.
This result is in sharp contrast to the assertion \cite{khan98} that 
a one-fluid theory is only possible for a cold plasma.  The charge-current 
tensor approach produces a fourth ``energy'' equation for the 
beamed current that can take into account fast current drift velocities and
hot plasma as well as fast bulk speeds. 

While of course valid in more benign environments, the charge-dynamical
form of Ohm's law probably will be needed only in very violent environments
such as electromagnetic, rotating black hole {\em formation}. These events can
have significant fluid and metric shear that may affect charge flow on
time scales shorter than the current rise time.  However, it is precisely
these environments that are believed to obtain in ``collapsars'', which
form in the centers of massive stars, and in mergers of neutron stars
in close binary systems.  And it is these systems that are believed to
lead directly to the formation of electromagnetic jets and their 
associated observable events.  Charge dynamics, therefore, may play
a significant role in understanding the engine that generates the highly
relativistic jets seen in gamma-ray bursts and other sources.

\begin{acknowledgments}
The author is grateful for discussions with M. Miller. 
He is also grateful for a JPL Institutional Research and Development
grant, and for the continued hospitality of the TAPIR group at Caltech.
The research described in this paper was carried out at the Jet Propulsion 
Laboratory, California Institute of Technology, under contract to the 
National Aeronautics and Space Administration.
\end{acknowledgments}

\begin{appendix}{}
\section{Velocity Integrals in Three-Space}
\label{three_d_integrals}

The velocity integrals in sections \ref{multifl} and \ref{onefl} were
cast as ones over the four-dimensional volume $\hat{\Omega}_u$. This causes
each integral, even those for quantities that appear in nonrelativistic
dynamics ($n_a$, $p_a$, $\varepsilon_a$), to contain a Lorentz factor
(equation \ref{lorentz_factor}) in its kernel.  In this Appendix we
show that each can be converted to its more familiar, three-space form
by confining the integration to take place on the mass hyperboloid only.

We begin by defining a three-dimensional density 
\begin{eqnarray}
\label{density6}
f_a & \equiv & 
\frac{\partial^6 {\rm N}_a} {m_a^3 \; \partial \Upsilon_u \; \partial \Upsilon_x}
\; = \; \frac{\partial^6 {\rm N}_a} {\partial \Upsilon_p \; \partial \Upsilon_x}
\end{eqnarray}
as the distribution of particles in three-momentum ${\bm p}$, where $m_a$ 
is the particle rest mass and $d \Upsilon_p \; = \; m_a^3 \, d \Upsilon_u$. 
This distribution is related to $\aleph_a$ by a delta function that enforces 
the mass hyperboloid
\begin{eqnarray}
\aleph_a & = & m_a^3 \, f_a \; \delta( \epsilon - \gamma^2 c/2) 
\; = \; m_a^3 \, f_a \; \frac{\delta( {\rm u}^{\hat{0}} - \gamma \, c ) }
{{\rm u}^{\hat{0}} / c}
\end{eqnarray}
(See equation \ref{lor_u_vol_elm}.)

The four-volume integral of any kernel $K$, weighted by $\gamma$, now can be 
replaced by a three-integral over $f_a$ on the mass shell, with no $\gamma$ 
weighting:
\begin{eqnarray}
\int_{\hat{\Omega}_u} \gamma \, \aleph_a \, K( {\bf x}, {\bf u} ) \, d \hat{\Omega}_u 
& = & \int_{\Upsilon_u} \int_{{\rm u}^{\hat{0}}} \gamma \, m_a^3 \, f_a \, 
\frac{\delta( {\rm u}^{\hat{0}} - \gamma \, c ) }{{\rm u}^{\hat{0}} / c}
\, K( {\bf x}, {\rm u}^{\hat{0}}, {\bm p} ) 
\, d \hat{\Omega}_u 
\nonumber
\\
& = &  \int_{\Upsilon_p} f_a \, d^3 p 
\; \int_{{\rm u}^{\hat{0}}} \frac{\gamma}{{\rm u}^{\hat{0}}/c} \, \delta( {\rm u}^{\hat{0}} - \gamma \, c ) 
\, K( {\bf x}, {\rm u}^{\hat{0}}, {\bm p} ) \, d {\rm u}^{\hat{0}} 
\nonumber
\\
& = & \int_{\Upsilon_p} K( {\bf x}, \gamma \, c, {\bm p} ) \, f_a \, d^3 p
\end{eqnarray}

Equations (\ref{spec_dens}), (\ref{spec_vel}), (\ref{hyd_energy}), and
(\ref{hyd_pressure}) now take on their familiar forms
\begin{eqnarray*}
n_a & = & \int_{\Upsilon_p} f_a \, d^3p
\\
{\bf V}_a & = & \frac{1}{n_a} \, \int_{\Upsilon_p} {\bf v} \, f_a \, d^3p
\\
\varepsilon_a & = & \int_{\Upsilon_p} (\gamma - 1) \, m_a \, c^2 \, f_a \, d^3p
\\
p_a & = & \frac{1}{3} \, \int_{\Upsilon_p} {\bm p} \cdot {\bf v} \, f_a \, d^3p
\end{eqnarray*}
(Recall that in the local Lorentz frame of the fluid ${\bf
v} = (0,~v^1,~v^2,v^3)$.)  However, ${{\bf V}'}_a$ in equation
(\ref{enh_spec_vel}), which is used to construct the beamed current ${\bf
j}'$, has no nonrelativistic analog (other than ${\bf j}$ itself when
$|{\bf v}| << c$), and therefore must always involve the Lorentz factor
\begin{eqnarray*}
{{\bf V}'}_a & = & \frac{1}{n_a} \, \int_{\Upsilon_p} \gamma \, {\bf v} \, f_a \, d^3p
\end{eqnarray*}
% (Note that an integral is not needed for $n_a'$, because it splits into the 
% known quantities $n_a$ and $\varepsilon_a$.)

\end{appendix}

\clearpage


%% If you are not including electronic art with your submission, you may
%% mark up your captions using the \figcaption command. See the 
%% No more than seven \figcaption commands are allowed per page, 
%% so if you have more than seven captions, insert a \clearpage 


\begin{thebibliography}{}
\bibitem[Anandan 1984]{anan84} Anandan, J. 1984, Class Quantum Grav, 1, 
    L51--6.
\bibitem[Andr\'{e}asson 2002]{and02} Andr\'{e}asson, H. 2002, Living Rev. 
    Relativity 5, 7. [Online article]: http://www.livingreviews.org/lrr-2002-7.
\bibitem[Ardavan 1976]{ard76} Ardavan, H. 1976, \apj, 203, 226--232.
\bibitem[Blackman \& Field 1993]{bf93} Blackman, E. G. \& Field, G. B. 1993, 
    Phys. Ret. Lett., 71, 3841--4.
\bibitem[Duncan \& Hughes 1994]{dh94} Duncan, G. C. \& Hughes, P. A. 1994, 
    \apj, 436, L119--122.
\bibitem[Eckart 1940]{eck40} Eckart, C. 1940, Phys. Rev., 58, 919--924.
\bibitem[Gedalin 1996]{ged96} Gedalin, M. 1996, Phys. Rev. Lett., 76, 
    3340--3.
\bibitem[Hughes {\it et al.} 2002]{hmd02} Hughes, P. A., Miller, M. A., 
    \& Duncan, G. C. 2002, \apj, 572, 713--728.
\bibitem[Krall \& Trivelpiece 1973]{kt73} Krall, N. A. \& Trivelpiece, A. W. 
    1973, Principles of Plasma Physics, (New York: McGraw-Hill). 
\bibitem[Khanna 1998]{khan98} Khanna, R. 1998, \mnras, 294, 673--681.
\bibitem[Koide 2003]{k03} Koide, S. 2003, Phys. Rev. D, 67, 4010--4024.
\bibitem[Lee {\it et al.} 2001]{lee01} Lee, H.K., Lee, C.H., \& van Putten, 
    M.H.P.M. 2001, \mnras, 324, 781--784.
\bibitem[Lichnerowicz 1967]{lich67} Lichnerowicz, A. 1967, Relativistic 
    Hydrodynamics and Magnetohydrodynamics, (New York: Benjamin). 
\bibitem[Melatos \& Melrose 1996]{mm96} Melatos, A. \& Melrose, D. B. 1996, 
    \mnras, 279, 1168--1190.
\bibitem[Marti \& Mueller 1999]{mm99} Marti, J. M. \& Mueller, E. 1999, 
    Living Rev. Relativity 2, 3. [Online article]: 
    http://www.livingreviews.org/lrr-1999-3.
\bibitem[Misner {\it et al.} 1973]{mtw73} Misner, C. W., Thorne, K. S., 
    \& Wheeler, J. A. 1973, Gravitation, (San Francisco: Freeman).
\bibitem[Punsly 2003]{puns03} Punsly, B. 2003, \apj, 538, 842--852.
\bibitem[Rossi \& Olbert 1970]{ro70} Rossi, B. \& Olbert, S. 1970, Introduction 
    to the Physics of Space, (New York: McGraw-Hill). 
\bibitem[Wald 1974]{wald74} Wald, R.M., Phys. Rev., D10, 1680.
\end{thebibliography}
\end{document}